\documentclass[aip, rsi, showpacs, amsmath, amssymb, reprint]{revtex4-1}
\usepackage{epsfig, graphicx, amsfonts, srcltx}
\usepackage{bm, hyperref, upgreek}
\pdfoutput=1

\begin{document}
\title{Scanning superconducting quantum interference device on a tip for magnetic imaging of nanoscale phenomena}

\author{A. Finkler}
  \email{amit.finkler@weizmann.ac.il}
\author{D. Vasyukov}
\author{Y. Segev}
\author{L. Ne'eman}
\author{E. O. Lachman}
\author{M. L. Rappaport}
\author{Y. Myasoedov}
\author{E. Zeldov}
\affiliation{Department of Condensed Matter Physics, Weizmann Institute of Science, Rehovot 76100,~Israel}
\author{M. E. Huber}
\affiliation{Department of Physics, University of Colorado, Denver,~CO~80217,~USA}

\date{\today}

\begin{abstract}
We describe a new type of scanning probe microscope based on a superconducting quantum interference device (SQUID) that resides on the apex of a sharp tip. The SQUID-on-tip is glued to a quartz tuning fork which allows scanning at a tip-sample separation of a few nm. The magnetic flux sensitivity of the SQUID is 1.8 $\mu\Phi_0/\sqrt{\mathrm{Hz}}$ and the spatial resolution is about 200 nm, which can be further improved. This combination of high sensitivity, spatial resolution, bandwidth, and the very close proximity to the sample provides a powerful tool for study of dynamic magnetic phenomena on the nanoscale. The potential of the SQUID-on-tip microscope is demonstrated by imaging of the vortex lattice and of the local AC magnetic response in superconductors.
\end{abstract}

\maketitle

\section{Introduction}
Within the rapidly developing field of nanotechnology there is a growing need for highly sensitive imaging techniques of local magnetic fields on the nanoscale \cite{Bending-1999, Kirtley2010, Degen2009, Gierling2011, Grinolds2011, Balasubramanian2008}. The magnetic force microscopy \cite{Moser-1995, Schwarz2008} and Lorentz microscopy \cite{Tonomura-1999} offer very high spatial resolution but lack in their field sensitivity. Conversely, scanning superconducting quantum interference device (SQUID) microscopes have the highest field sensitivity, but they currently suffer from rather low spatial resolution \cite{Kirtley2010}. In order to overcome this limitation, in recent years there is a growing interest in micro and nanoSQUIDs \cite{Foley2009, CleuziouJ.-P.2006, koshnick:243101, Nagel2011, lam:1078, PhysRevB.70.214513, hao:092504, Nb_nanoSQUID, Lam-2011, Granata2008}. Since the magnetic field of a nanostructure decays rapidly with distance, the measured spatial resolution is ultimately limited by the separation between the sample and the probe. In order to achieve nm scale resolution, the probe has to be able to approach and scan the sample within a few nm of the surface. One of the obstacles towards this goal is the location of the SQUID itself, i.e., on a wafer or chip, making it difficult to bring the sensor (or the pick-up loop, if separate from the sensor) closer than a few hundred nanometers to the sample. The main challenge is therefore to develop magnetic field sensors that have high field sensitivity, very small dimensions, and a compatible geometry. \\
\noindent In this paper we describe a new scanning probe microscope (SPM). The innovative element of the instrument is a nanoSQUID that resides on a sharp tip \cite{Finkler2010}, which is ideally suited for scanning microscopy. By attaching the SQUID-on-tip (SOT) to a quartz tuning fork (TF) we can scan the sensor within few nm from the surface of the sample, thus providing a quantitative, high sensitivity, and high spatial resolution tool for imaging and investigation of static and dynamic magnetic phenomena on the nanoscale. The feasibility of the method is demonstrated by imaging of the vortex lattice and of the local DC and AC magnetic response in Al, NbSe$_2$ and Nb superconductors. In addition, we present nanoscale topography imaging capability of the microscope along with the magnetic imaging.

\section{Instrumentation}
 \label{Instr}
  \subsection{SQUID-on-tip}
    \subsubsection{Fabrication}
\begin{figure}[ht!]
       \includegraphics[width = 0.475\textwidth]{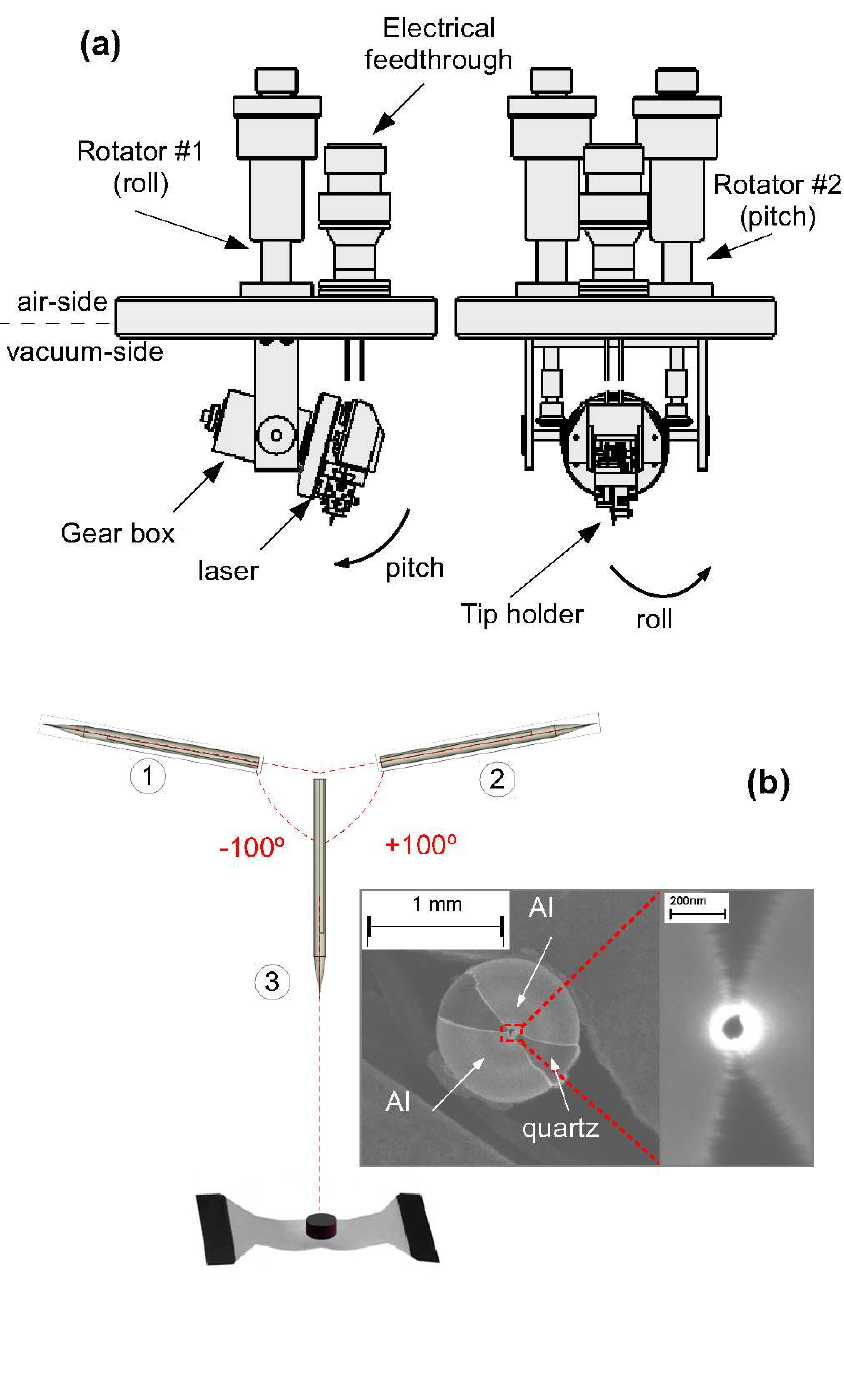}
       \caption{(a) A drawing of the rotator used to hold and rotate the tip during deposition based on the design of Yoo et al. \cite{Yoo1997} (b) A schematic of the three step thermal deposition of a SOT. The tip in each step is offset for clarity from its actual position above the source. Inset: a SEM micrograph showing a head-on view of the SOT. The darker regions are  quartz covered with not more than 3 nm of aluminum, below its percolation threshold, and the brighter regions are the deposited aluminum. The dark spot in the center surrounded by the bright ring is the hole in the SOT loop.}
       \label{deposition}
      \end{figure}

      Using a commercial pipette puller \cite{Note1}, we pull a quartz tube with 1~mm outer diameter and an inner diameter of 0.5~mm to form a pair of sharp pipettes with a tip diameter that can be controllably varied between 100 and 400~nm. Then, we either solder a thin layer of indium or evaporate a 200~nm-thick film of gold on two sides of the cylindrical base of the pipette. Afterwards, the pipette is mounted on a rotator and put into a vacuum chamber for three steps of thermal evaporation of aluminum. The rotator (see Fig.\ \ref{deposition}a) has an electrical feedthrough for two operations. First, a 2~mW red laser diode is mounted colinear with the tip, which is used to point the tip towards the very center of the source. This defines the zero angle. The second electrical connection is for an \textit{in situ} measurement of the tip's resistance during deposition. In the first step, 25~nm of aluminum are deposited on the tip tilted at an angle of -100$^\circ$. Then the tip is rotated to an angle of +100$^\circ$ for a second deposition of 25~nm, as shown in Fig.\ \ref{deposition}b.
As a result, two leads are formed on opposite sides of the quartz
tube separated by gap regions of bare quartz. In the last step,
20-22~nm of aluminum are deposited at an angle of 0$^\circ$,
coating the apex ring of the tip. Due to geometrical
considerations, in this step about 3 nm of aluminum are also
deposited in the gap regions at a very shallow deposition angle.
This thin layer, however, does not percolate and hence does not
short the two leads. The resulting structure is therefore a ring
connected to two leads (see inset in Fig.\ \ref{deposition}b).
``Strong'' superconducting regions are formed in the areas where
the leads make contact with the ring, while the two parts of the
ring in the region of the gap between the leads constitute two
weak links, thus forming the SQUID. A finite resistance appears
after depositing approximately 7 nm in the third step. The typical
final room-temperature resistance of the SOT is 1.5~k$\Omega$.
Since the device is highly sensitive to electrostatic discharge
(ESD), we follow a strict protocol of anti-ESD procedures, which
include, among others, the use of grounding strips, mats and
physical electrical shorts on the tip during transfer.
    \subsubsection{Characterization}
\begin{figure}
       \includegraphics[width = 0.475\textwidth]{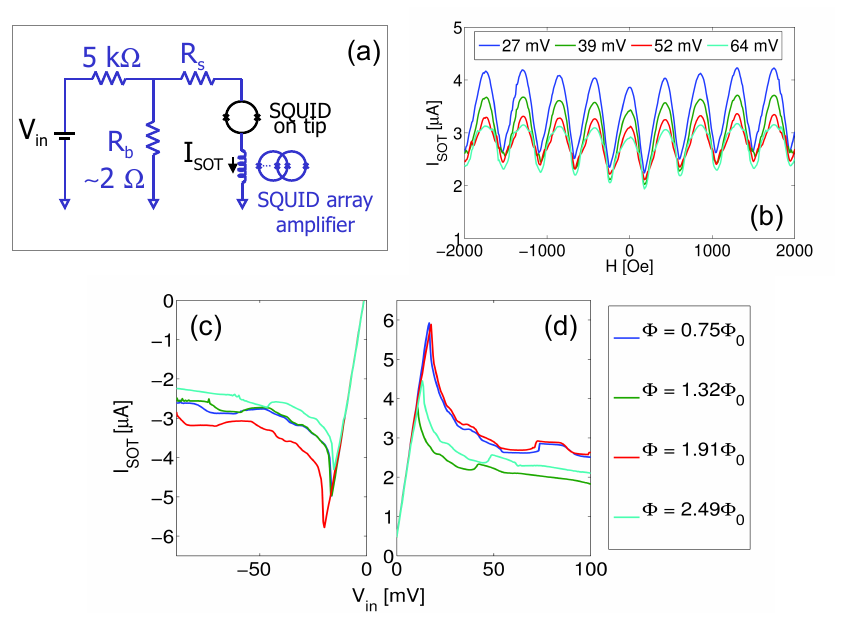}
       \caption{ (color online) (a) Schematic of the measurement circuit. A room-temperature voltage source in series with a cold 5 k$\Omega$ resistor applies a nearly constant current to the SOT shunted by a small resistance, $R_b$. When the SOT is in the superconducting state, the entire current passes through it and is measured via inductive coupling by the SSAA. When it switches to the voltage state, most of the current flows through $R_b$, while a small fraction continues to flow through the SOT itself. $R_s$ is a small, unwanted, series resistance, typically less than 1 $\Omega$. (b) $I_\text{SOT}$ vs.\ $H$ for different values of $V_\text{in}$ showing the quantum oscillations of the current with a period that corresponds to an SOT loop diameter of 245 nm. (c)-(d): $I_\text{SOT}$ vs.\ $V_\text{in}$ for different values of applied magnetic field, $H$. The field is shown in terms of the flux in the SOT in units of the flux quantum, $\Phi_0$.}
       \label{circuit}
      \end{figure}

      A series SQUID array amplifier\cite{Huber-2001} (SSAA) is used as a cryogenic current-to-voltage converter to read out the SOT. The SOT is connected in a voltage-bias configuration, as shown in Fig.\ \ref{circuit}a.
      For each value of the applied magnetic field, $H$, we measure the current through the SOT, $I_\text{SOT}$, as a function of the input voltage, $V_\text{in}$. At low $V_\text{in}$, the SOT is in its superconducting state, and $I_\text{SOT}$ is linear with $V_\text{in}$. At $V_\text{in} > 25$ mV, the SOT switches to its voltage state and a decrease in the current through it is observed, plotted in Figs.\ \ref{circuit}c and \ref{circuit}d. At higher values of $V_\text{in}$, the SOT behaves again as a resistive element. Similary, Fig.\ \ref{circuit}b shows the SOT response upon sweeping the applied field $H$ at various constant biases. After mapping the entire $I_\text{SOT}\left(V_\text{in}, H\right)$ space and measuring the current noise in each point in that space, we choose a working point of $V_\text{in}$ and $H$, at which the ratio between the noise (in units of A/Hz$^{1/2}$) and the sensitivity, $dI_\text{SOT}/dH$ (in units of A/Oe) is minimal. Since the response is periodic in field, and since our SOTs are usually asymmetric \cite{Finkler2010} with respect to applied field and polarity of $V_\text{in}$, there is a wide range of fields and polarities at which high sensitivity can be achieved. After choosing a working point, we perform a calibration measurement to determine the exact G/A converstion ratio.
  \subsection{Tuning-fork microscopy}
    \subsubsection{Assembly}
\begin{figure}[ht!]
       \includegraphics[width = 0.475\textwidth]{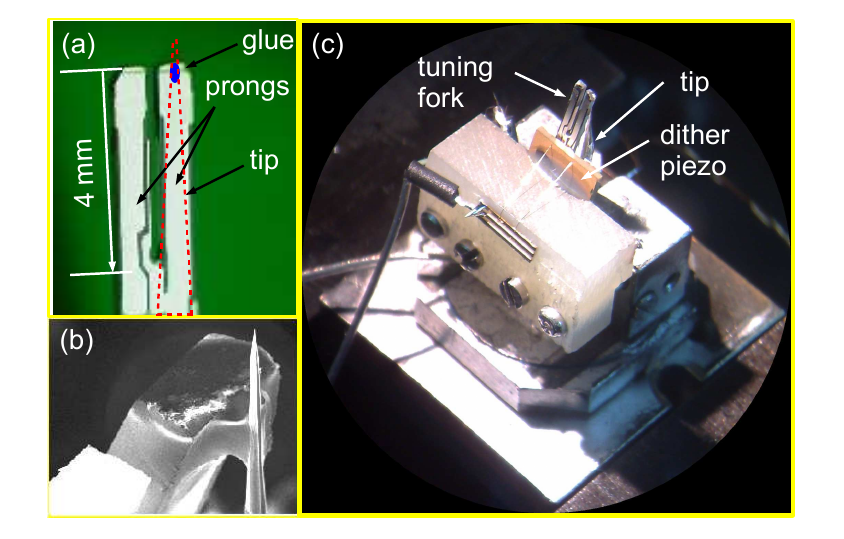}
     \caption[SEM of a tuning fork assembly]{(color online) (a) Image of a tuning fork after the vacuum can had been removed. The outline of the tip is superimposed on the right prong in a red dashed line. The blue blob near the end of the prong shows the location and approximate size of the epoxy drop used to glue the tuning fork to the tip; (b) A SEM micrograph of a tip glued to a prong of a tuning fork; (c) The tip/tuning fork assembly. The tip is positioned between two stainless steel blocks, one with a groove, acting as an electrode, and the other having a BeCu spring connected to it, acting as a counter-electrode by pressing against the tip. The tuning fork itself is glued onto the dither piezo with Bondmaster E32 and then glued to the tip with Araldite 2020.}
     \label{tuning_fork_assembly}
      \end{figure}

      Commercial quartz tuning forks (TF), normally used as time bases in digital watches, are employed as force sensors. They are laser-trimmed to have a resonance at $2^{15}$ Hz, and typically have a high Q-factor due to low internal losses (dissipation) in quartz. We glue the SOT to one prong of the TF, whose length is 4 mm, width 0.6 mm and thickness 0.34 mm (Figs.\ \ref{tuning_fork_assembly}a and \ref{tuning_fork_assembly}b). When the SOT approaches the surface of the sample, electrostatic and van der Waals forces dampen the TF's resonance and shift its frequency\cite{Giessibl}. Tuning fork microscopy uses either the decrease in amplitude or the change in frequency as the feedback parameter. In our setup we monitor the frequency shift of the TF using a phase-locked loop (PLL). We can detect a small frequency increase at about 20 nm from the surface that grows on approaching the surface. At 10 nm from the sample surface the increase of the TF resonant frequency is typically about 100 mHz. Preparing a TF for microscopy entails the removal of its vacuum can and the two wires soldered to its contacts. It is then glued to a 5$\times$5$\times$0.5 \text{mm}$^3$ quartz piece which was pre-coated with two gold pads (70 \AA\ Cr/2000 \AA\ Au). Alternatively, this quartz piece is replaced by a dither piezo having the same dimensions (shown in Fig.\ \ref{tuning_fork_assembly}c). The two TF contacts are wire-bonded to those gold pads, which are in turn connected to external wires via two additional wire bonds.

      One can electrically excite the TF with a voltage source and read the current through it using a current-to-voltage converter \cite{grober:2776}. However, it is more advantageous to decouple the excitation from the readout by using the voltage source to drive the dither piezo, which mechanically excites the TF. This latter method reduces the effect of stray capacitances of the cables and the electrical contacts themselves from entering the readout signal. We used a shear-dither piezo \cite{Note2} for this purpose, which was used both as the plate to which we glued the TF and as its mechanical excitation medium (shown in Fig.\ \ref{tuning_fork_assembly}c).
    \subsubsection{Tip-sample interaction and electronics}
\begin{figure}[ht!]
       \includegraphics[width = 0.4\textwidth]{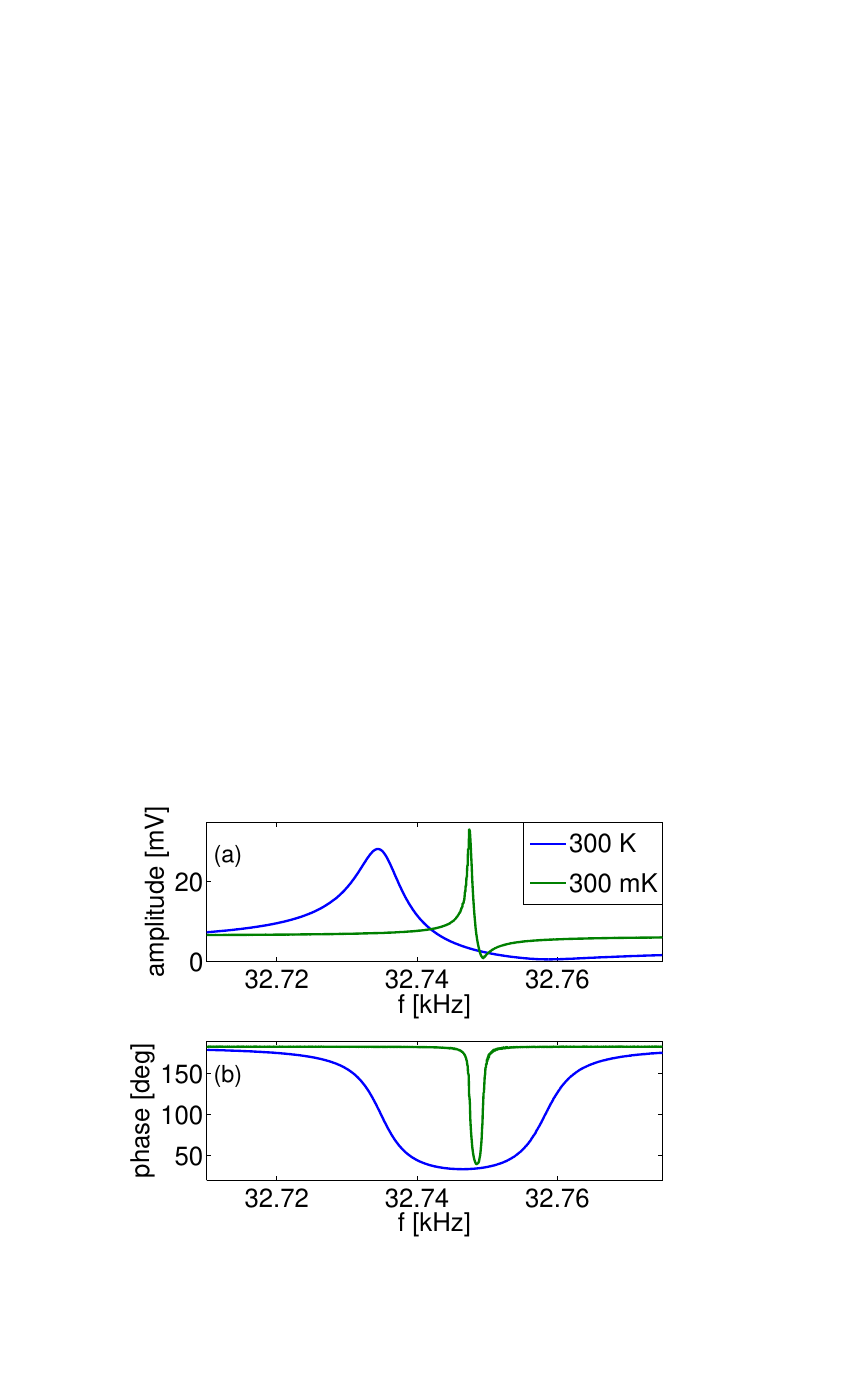}
       \caption{A comparison between room temperature, atmospheric pressure and low temperature (300 mK), low pressure resonance curves of the tuning fork. The room temperature resonance is typically wider and its resonant frequency is slightly lower than the low temperature one.}
       \label{resonant_curves}
      \end{figure}

      In our microscope setup, as shown in Fig.\ \ref{tuning_fork_assembly}c, the tip is positioned between two stainless steel electrodes, such that each SOT lead is in contact with one of the two. External wires are soldered to the electrodes to provide electrical contact to the SOT. We glue the SOT to one of the prongs of the TF. This dampens the resonance \cite{Shelimov2000} dramatically. Therefore, we try to use as small an amount of glue as possible. Using a two-part epoxy with an extremely low viscosity (150 mPa$\cdot$s) enables us to place a very small drop of glue (Fig.\ \ref{tuning_fork_assembly}b). The farther the SOT protrudes above the edge of the prong, the smaller its effective spring constant \cite{Karrai2000} becomes (for a cylindrical tip):
      $$
    k_\text{eff} = \frac{3\pi E r^4}{4l^3},
      $$
      where $r$, $l$ and $E$ are the radius, length and Young modulus of the tip, respectively.
      Thus, if the SOT protrudes too much, the electrostatic forces of interaction with the sample will dampen its motion while the tuning fork's prong remains almost unaffected. Consequently, we always try to position the tip with respect to the tuning fork so that the SOT itself protrudes only slightly above the edge of the prong (typically a few tens of microns). We then excite the tuning fork and measure its resonance, specifically noting the location of the peak, its amplitude and its width. This measurement is performed at room temperature and atmospheric pressure and repeated at low temperature (300 mK) in vacuum for comparison. Typical resonance curves for the TF at room temperature and low temperature are plotted in Fig.\ \ref{resonant_curves}.
      More often than not, the low temperature, high vacuum resonance of the tuning fork is much sharper than at room temperature and atmospheric pressure. One can monitor the height of the resonance peak and set the feedback threshold to, for example, 50\% of the maximum. This method, also known as slope detection \cite{Albrecht}, is not fast enough, since quartz tuning forks have relatively high Q-factors, especially at low temperatures (typically a few 100,000 for a bare TF and a few 10,000 for a TF glued to a tip) \cite{Albrecht, durig:3641}. These high Q-factors result in a long settling time, of the order of seconds, which is unacceptable for scanning probe microscopy. We have used a phase-locked loop (PLL, either RHK's PLLpro or attocube's ASC500) to increase the bandwidth and scanning speed \cite{Albrecht, Gildemeister}. Interaction forces between the tip and the sample result in a frequency shift of the tuning fork's resonance. The basic idea behind the PLL is to track this resonance in a closed loop circuit and compare its current location with the fundamental one to give the frequency shift, $\Delta f$. This frequency shift serves as the error function of the height, or $z$, feedback loop and hence allows simultaneous imaging of the magnetic field and sample topography. In our setup, the typical scanning speed was between 0.5-1 $\upmu\mathrm{m}$/sec when using the PLL.
  \subsection{Microscope design}
      The microscope assembly, described below, is mounted on a $^3$He Oxford HelioxTL probe, which was modified for this purpose. It is designed to work in vacuum at a base temperature of less than 300 mK. All crucial signals (SSAA, tuning fork, SOT) are transmitted via coaxial cables, while the rest are passed through either Manganin or NbTi twisted-pairs in order to reduce the heat load from room temperature.
    \subsubsection{Schematics of the assembly}
\begin{figure}[ht!]
       \includegraphics[width = 0.475\textwidth]{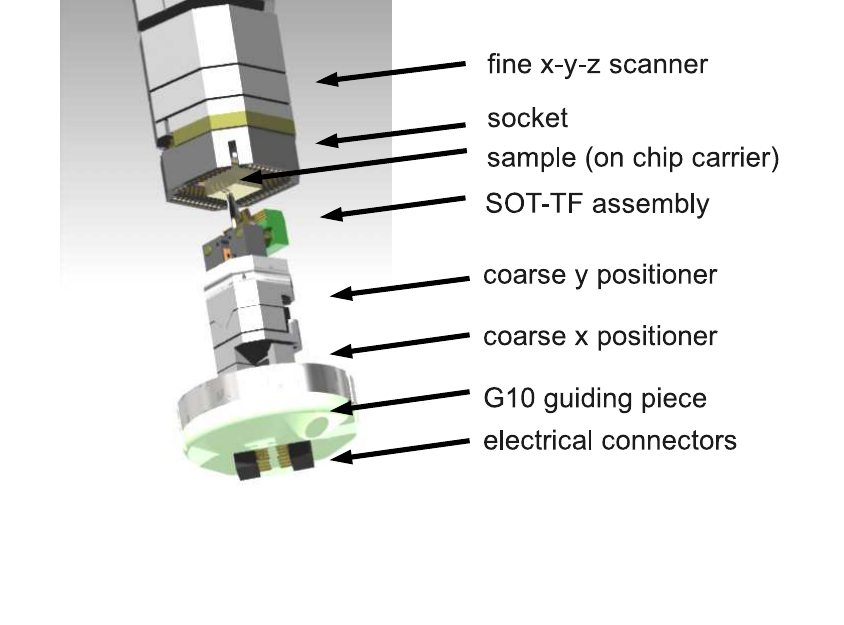}
       \caption[Schematics of the microscope assembly]{The microscope assembly without its outer shell. The lower part includes the coarse $x$ and $y$ positioners and the tip holder while the upper part includes the $z$ positioner, the $xyz$ scanners and the sample holder.}
       \label{microscope}
      \end{figure}

      The coarse motion is performed by three attocube \cite{Note3} positioners (1$\times$ANPz101RES, 2$\times$ANPx50), one for each axis, based on the slip-stick mechanism. The positioners themselves are driven by a saw tooth pulse from a high-voltage driver (attocube ANC-150). The z-axis positioner incorporates a resistive encoder that gives its absolute position to an accuracy of 2 $\upmu$m and resolution of 200 nm. Scanning is performed by an attocube integrated xyz scanner (ANSxyz100), with an x-y scan range of $30 \times 30\ \upmu\mathrm{m}^2$ and a z-scan range of 15 $\upmu\mathrm{m}$, all at $T=300$ mK. The microscope is shown in Fig.~\ref{microscope}.
       We drive these piezoelectric scanners with voltage from an SPM controller (either RHK SPM-100 or attocube ASC500). As mentioned above, the output of the PLL (in the form of a frequency shift, $\Delta f$) serves as the input of the z-height feedback loop. This feedback loop controls the height of the tip above the sample by changing the voltage applied to the z-scanner.
    \subsubsection{Vibration/isolation and acoustics considerations}
      The cryostat was isolated from floor vibrations by suspending it in the center of a hollow marble block supported by four commercial isolators \cite{Note4}; the weight of the marble is 920 kg. These isolators attenuate vibrations from the environment at low frequencies, (-20 dB at 10 Hz). The disadvantage of these isolators is that they have a resonance at around 1 Hz. To reduce acoustical noise, we wrapped the dewar with an acoustic blanket \cite{Note5} that attenuated the vibrations of the dewar's outer vacuum chamber by 3 dB at its resonance frequency of 186 Hz. In addition, the top of the cryostat was covered by an acoustic enclosure \cite{Note6} that sits on the marble and seals well to its smooth surface. The enclosure provides further attenuation of 45 dB at 500 Hz. The combination of the isolators and the enclosure created a sufficiently quiet environment in which we were able to safely scan samples with the tip-sample separation of a few nanometers.
\section{Measurements}
    With its magnetic sensor coupled to a topography sensor, the scanning SQUID microscope can acquire simultaneously both the topography and magnetic signals of the sample. Below we present measurements of three different types of samples, exhibiting the potential of the instrument.
 \label{Meas}
  \subsection{Self-induced magnetic field of a transport current}
\begin{figure}
     \centering
     \includegraphics[width = 0.475\textwidth]{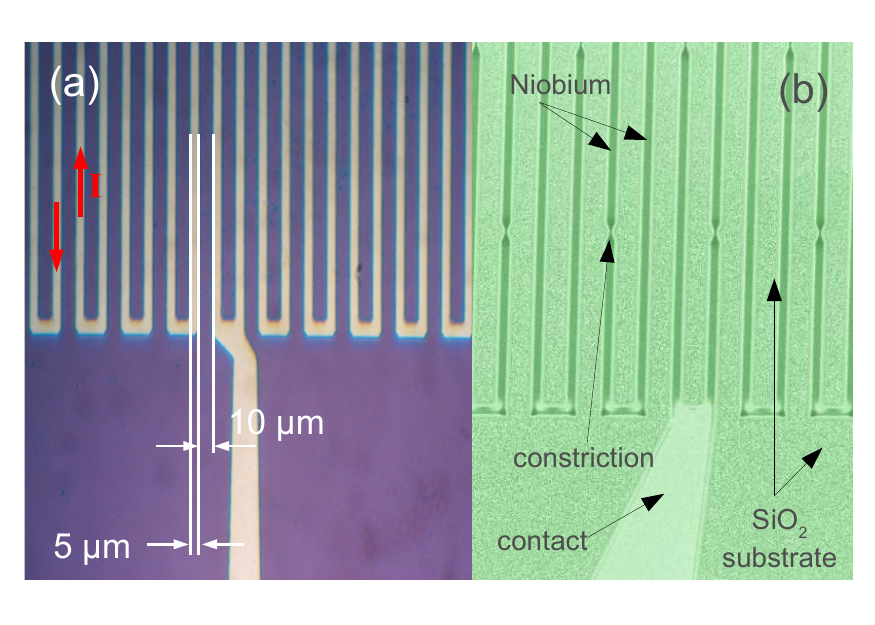}
     \caption[Serpentine samples]{(a) An optical image of the Al serpentine deposited on a Si/SiO$_2$ substrate. The arrows indicate the direction of the current flow in the serpentine. (b) A~SEM image of a similar serpentine made of Nb that contains periodic constrictions (narrowing in the line shape).}
     \label{al_meander}
    \end{figure}

\begin{figure}
         \includegraphics[width = 0.475\textwidth]{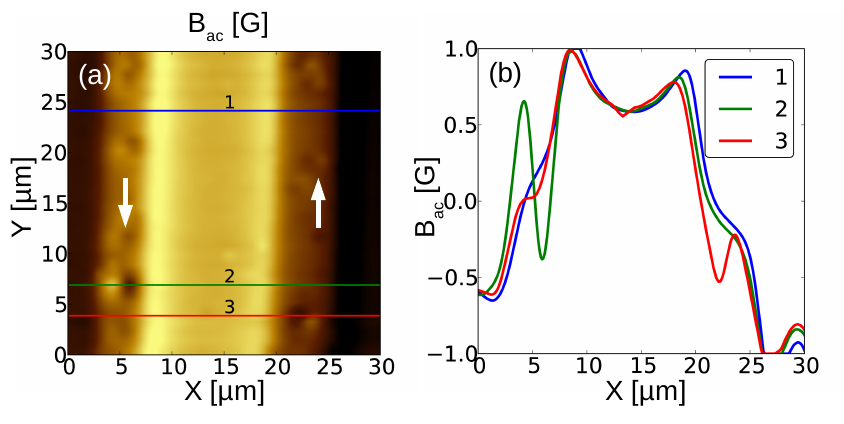}
         \caption{AC magnetic field $B_{ac}$ at a height of 1 $\upmu\mathrm{m}$ above an aluminum serpentine carrying an AC current of 2 mA at 510 Hz under an applied DC field of 90 Oe. (a) A 30$\times$30 $\upmu\mathrm{m}^2$ magnetic image showing two strips of the serpentine carrying current in opposite directions as indicated by the arrows. The circular ``depressions'' along the central axes of the strips are normal (not superconducting) regions in the intermediate state of type-I superconductor. The applied AC current circumvents the normal regions giving rise to the observed ``depressions'' in the local $B_{ac}$. (b) Line scans from (a), with lines '2' and '3' passing through ``depressions''.}
         \label{biot-savart}
    \end{figure}

    Our first calibration sample was an aluminum serpentine, 200 nm thick, with a line width of 5 $\upmu\mathrm{m}$ and a period of 15 $\upmu\mathrm{m}$ (see Fig.\ \ref{al_meander}a).
    Already at a distance of 10 $\upmu$m from its surface, the SOT was able to image the AC magnetic field generated by an AC current of 2 mA in the serpentine. Figure \ref{biot-savart} shows the self-induced field due to two adjacent strips of the serpentine, measured in constant height mode without feedback on $z$. Following Biot-Savart law in superconducting strip geometry \cite{Zeldov1994}, the AC transport current in the left strip gives rise to a positive (in-phase) field on the right edge of the strip and a negative (out-of-phase) signal on the left edge. The adjacent strip, carrying current in the opposite direction, generates an inverted field profile. The resulting combined field is high in-between the strips (bright color) and low (dark) outside the strips. In addition, we resolve fine structure within the strips. Since our Al film is type-I superconductor, small normal regions are formed in the central part of the strips in the presence of applied DC field of 90 Oe. The applied transport current flows only in the superconducting regions, avoiding the normal areas, giving rise to the ``depressions'' in the local AC field.
  \subsection{Magnetic screening in a NbSe$_2$ crystal}
\begin{figure}
     \includegraphics[width = 0.475\textwidth]{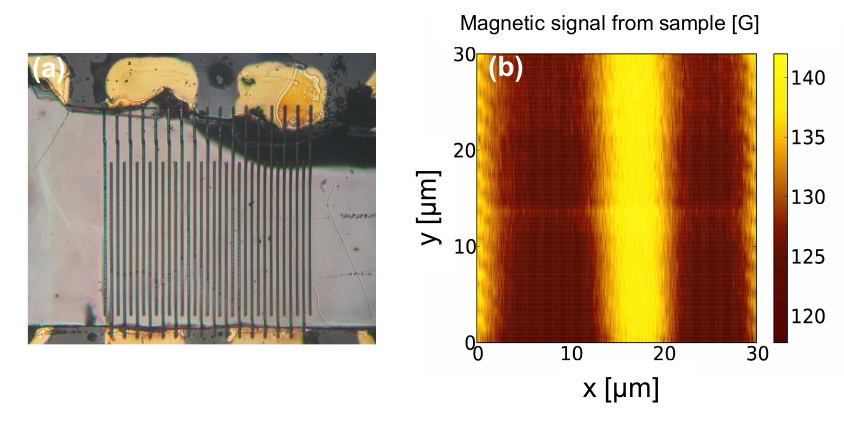}
     \caption{(a) Optical image of the NbSe$_2$ crystal patterned into a serpentine using a focused ion beam. (b) Image of the magnetic signal at tip-sample separation of 2.5 $\upmu$m exemplifying the screening of the magnetic field in the superconductor at applied field of 140 Oe. The dark regions are two NbSe$_2$ strips that partially screen the magnetic field and the bright strip is the FIB etched region between the two superconducting strips.}
     \label{NbSe2}
    \end{figure}

    Aluminum, being a type-I superconductor, does not exhibit vortices and therefore one cannot observe them in thick films of such a material. In order to investigate type-II superconductors we have prepared serpentines of NbSe$_2$ single crystal and of Nb thin films. A 10 $\upmu$m thick cleaved NbSe$_2$ crystal was milled using a focused ion beam (FIB) to a serpentine shape as shown in Fig.\ \ref{NbSe2}a. The sample was cooled in a low field and then the applied field was increased to 140 Oe. The local magnetic field was then imaged at a height of 2.5 $\upmu$m above the surface of the serpentine as shown in Fig.\ \ref{NbSe2}b. The screening currents partially shield the magnetic field in the NbSe$_2$ strips (dark) and enhance the field in-between the strips (bright).
  \subsection{Vortex matter in thin niobium films}
\begin{figure}[ht!]
     \includegraphics[width = 0.475\textwidth]{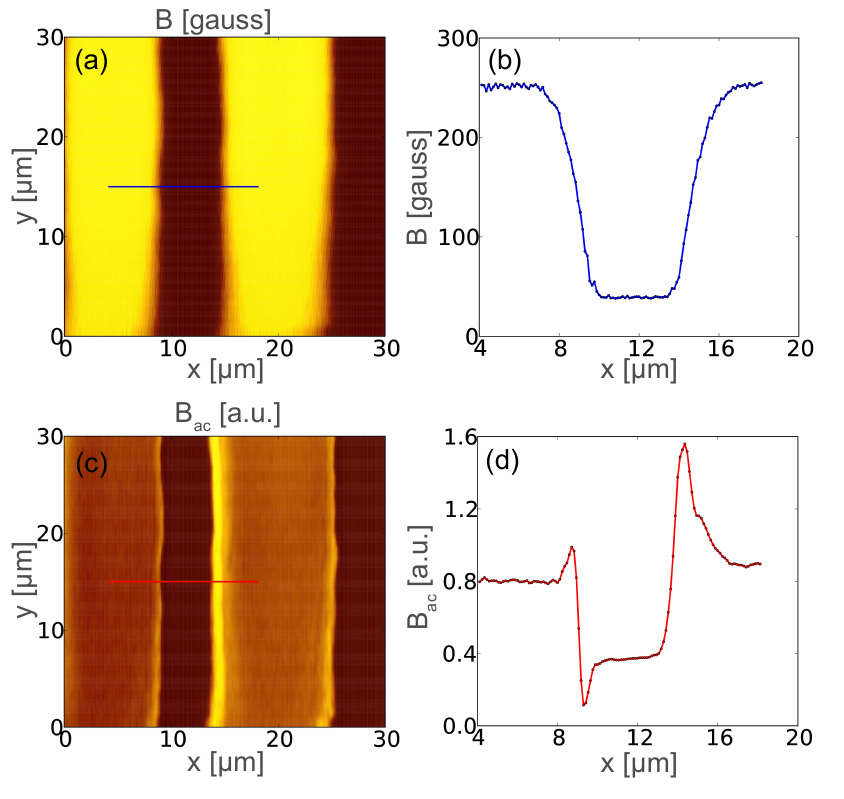}
     \caption{Meissner-like signal from a Nb serpentine at an applied field of 250 Oe.
(a) The induced field, $B_z$, scanned a few microns above the
sample. The dark (38 G) regions are the Nb strips while the bright
(250 G) regions are the silicon substrate. (b) A cut along the
$x$-axis, marked by the blue line in (a). The shielding currents
screen the superconducting Nb strips from the applied magnetic
field. The trapped field of 38 G in the Nb strip is due to
field-cooling in the presence of the remnant field in the magnet.
(c) AC signal measured by the SOT at 32 kHz due to the
oscillations of the TF and the SOT along $x$ axis. (d) A cross
section along the $x$-axis, marked by the red line in (c). Since
the measured AC signal is determined by the gradient of the DC
field, the spikes correspond to the sudden change in the local
field when crossing the edges of the superconducting strip. The
offset values of $B_{ac}$ apparently result from electrical pickup
at 32 kHz and a contribution from in-plane field $B_x$ because the
tip is not exactly perpendicular to the sample surface.}
     \label{dIdx}
    \end{figure}

    Nb film was deposited on a Si substrate using an e-gun while keeping the substrate at a temperature of 300 $^\circ$C in a background pressure of $10^{-6}$ Torr. The 200 nm-thick film was patterned into a serpentine structure as shown in Fig.\ \ref{al_meander}b. The sample was cooled in a small residual field, then the applied field was increased and the local magnetic field was imaged by the scanning SOT at 250 Oe, as shown in Figs. \ref{dIdx}a and  \ref{dIdx}b. The dark regions are the Nb strips that are essentially fully screened due to very strong vortex pinning in Nb films at low temperatures.

    In order to demonstrate operation of the SOT at high frequencies we take advantage of the fact that the TF and the tip oscillate at 32 kHz in the plane parallel to the sample surface. In presence of a field gradient $dB_z/dx$ this oscillation gives rise to an effective AC magnetic field as sensed by the SOT, $B_{ac} = \Delta x (dB_z/dx)$, where $\Delta x$ is the amplitude of the tip oscillation along the $x$ axis. We can vary $\Delta x$ from below a nm to few tens of nm. Using a lock-in amplifier we can thus measure simultaneously the DC and the AC magnetic signals while scanning the sample. Figure \ref{dIdx} shows an example of such simultaneous measurement of $B_z$ and $B_{ac}$ at 32 kHz in a Nb serpentine scanned a few microns above the sample in constant height mode, in a DC field of 250 Oe.

\begin{figure}[ht]
     \includegraphics[width = 0.475\textwidth]{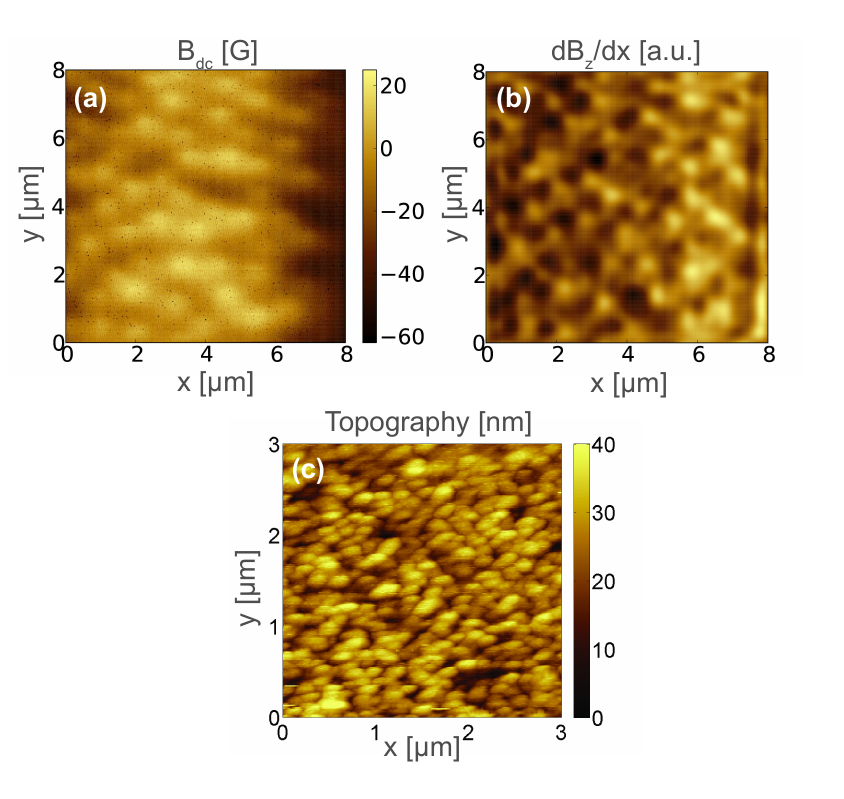}
     \caption{(a) The DC magnetic signal $B_z$ in Nb film measured after field-cooling the sample in 20$\pm 7$ Oe. Individual vortices are visible as bright spots. (b) The corresponding AC magnetic signal in the same scan area due to the oscillation of the SOT in the $x$ direction at 32 kHz. The vortices are visible as a pair of bright/dark signal corresponding to the gradient $dB_z/dx$ of the vortex field. (c) A topographic measurement of the surface of the Nb film, $3\times 3$ $\upmu\mathrm{m}^2$, in the same setup (different scan area).}
     \label{acdc_vortices}
    \end{figure}

    Finally, we demonstrate the high spatial resolution of the instrument by imaging vortices and surface topography in Nb film. For this the same sample was field-cooled in an applied magnetic field of 20$\pm 7$ Oe and the scanning was carried out in a closed feedback mode in which the tip maintains a constant height of few nm above the surface. Three signals are measured and imaged simultaneously during the scan: the DC SOT signal which provides $B_z$, the AC SOT signal at 32 kHz which reflects the $dB_z/dx$, and the $z$ feedback voltage which renders the sample topography. Figures \ref{acdc_vortices}a and \ref{acdc_vortices}b show vortices imaged in the Nb film by the DC and AC magnetic signals.
    The observed vortex lattice is highly disordered due to very strong pinning in Nb film at 300 mK (see Ref.\ \onlinecite{Brandt-1995} and references therein). In the DC image the vortices are visible as bright spots, while in the AC image the vortices appear as a bright-dark pairs from left to right as determined by the field gradient.  We can fit the spatial field dependence of the individual vortices to theoretical calculations \cite{Brandt2005} to find the corresponding magnetic penetration depth in the Nb film, which in this case turns out to be 400 nm. In the image area of $8\times 6\ \upmu\mathrm{m}^2$ there are about 34 individual vortices, corresponding to an average field of 14 G, in fair agreement with the applied field. Figure \ref{acdc_vortices}c shows a high resolution topographical image of the sample (in a different scan area), showing the granular structure of the Nb film. Most grains are 120 to 300 nm in diameter and 25 to 35 nm in height.

\section{Summary and future prospects}
 \label{future}
  We described a new scanning nanoSQUID microscope based on the development of a unique SQUID-on-tip. The unconventional geometry of the SQUID-on-tip together with the coupling to a quartz tuning fork allows for its assembly as a dual magnetic/topographic scanning probe microscope, making it the first SQUID acting as magnetic sensor in a scanning probe microscope with the sensor itself being only a few nanometers away from the surface of the sample. Since in most of the relevant microscopic systems the magnetic signal decays rapidly with the distance from the sample (distance cubed for magnetic dipoles, exponentially for a vortex lattice in superconductors, etc.), achievement of such a close sensor-sample separation is of major importance. The potential of the instrument has been demonstrated by study of DC and AC magnetic phenomena in type-I and type-II superconductors, including imaging the self-induced magnetic field of an AC transport current in Al thin films, observation of the magnetic shielding in NbSe$_2$ single crystal and in Al and Nb films, and imaging of disordered vortex lattice along with nanoscale surface topography in Nb films. In addition, the SQUID-on-tip is expected to become a very sensitive probe of magnetic dipoles due to its high flux sensitivity and very small size. Currently the projected \cite{Tilbrook2009} spin sensitivity of our Al SQUID-on-tip is 65 $\mu_B/\sqrt{\mathrm{Hz}}$ and there is a strong potential for development of more sensitive variations of the device towards the goal of single spin sensitivity. The scanning SQUID-on-tip microscope is thus a promising tool for study of a wide range of static and dynamic magnetic phenomena on the nanoscale.\\

  \noindent We acknowledge fruitful discussions with Shahal Ilani, Grigorii Mikitik, Ernst H. Brandt and Daniel Prober. This work was supported by the European Research Council (ERC) Advanced Grant, by the German-Israeli Foundation (GIF), and by the U.S.-Israel Binational Science Foundation (BSF).

\end{document}